\documentclass[twocolumn,amsmath,amssymb,prd,showpacs,superscriptaddress,preprintnumbers]{revtex4}
\usepackage{graphicx}
\usepackage{latexsym}
\usepackage{bm}
\newcommand{\beq}[1]{\begin{equation}\label{#1}}
\newcommand{\eeq}{\end{equation}}
\newcommand{\bea}[1]{\begin{eqnarray} \label{#1}}
\newcommand{\eea}{\end{eqnarray}}
\newcommand{\ba}{\begin{array}}
\newcommand{\ea}{\end{array}}

\newcommand{\rf}[1]{(\ref{#1})}
\def\be{\begin{equation}}
\def\ee{\end{equation}}
\def\gs{\mathrel{
   \rlap{\raise 0.511ex \hbox{$>$}}{\lower 0.511ex \hbox{$\sim$}}}}
\def\ls{\mathrel{
   \rlap{\raise 0.511ex \hbox{$<$}}{\lower 0.511ex \hbox{$\sim$}}}}
\newcommand{\D}{\displaystyle}

\newcommand{\unitmatrix}{\openone}

\newcommand{\Utbm}{U_{\rm TBM}}
\newcommand{\Ueps}{U_{\epsilon}}

\newcommand{\eps}{\epsilon}

\newcommand{\ra}{\rangle}
\newcommand{\la}{\langle}

\newcommand{\bad}{\begin{array}{ccc}}
\newcommand{\bav}{\begin{array}{cccc}}
\newcommand{\baf}{\begin{array}{ccccc}}

\def\U2{\underline{U\hspace{-.9mm}}\hspace{.9mm}}

\def\nua{{\nu_{\alpha}}}
\def\nub{{\nu_{\beta}}}

\def\numu{{\nu_{\mu}}}
\def\nutau{{\nu_{\tau}}}

\begin{document}

\title{Unitary (TriMinimal) Parametrization of Perturbations\\ to Tribimaximal Neutrino Mixing}

\author{Sandip Pakvasa}
\email[Electronic mail: ]{pakvasa@phys.hawaii.edu}
\affiliation{Department of Physics and Astronomy, 
University of Hawaii, Honolulu, HI 96822, USA}

\author{Werner Rodejohann}
\email[Electronic mail: ]{werner.rodejohann@mpi-hd.mpg.de}
\affiliation{Max--Planck--Institut f\"ur Kernphysik, 
Postfach 103980, D--69029 Heidelberg, Germany}

\author{Thomas J. Weiler}
\email[Electronic mail: ]{tom.weiler@vanderbilt.edu}
\affiliation{Department of Physics and Astronomy,
Vanderbilt University, Nashville, TN 37235, USA}

\date{\today}

\begin{abstract} 
Current experimental data on neutrino mixing are very well described by 
TriBiMaximal mixing. 
Accordingly, any phenomenological parametrization of the MNSP matrix must 
build upon TriBiMaximal mixing. 
We propose one particularly natural  parametrization,
which we call ``TriMinimal''.  
The three small deviations of the PDG angles from their TriBiMaximal values,
and the PDG phase, parametrize the TriMinimal mixing matrix.
As an important example of the utility of this new parametrization,
we present the simple resulting expressions for the flavor-mixing probabilities of 
atmospheric and astrophysical neutrinos.
As no foreseeable experiment will be sensitive to more than second order in the small parameters,
we expand these flavor probabilities to second order.

\end{abstract}


\maketitle

Neutrino physics has entered the precision era \cite{reviews}. In the next decade 
the uncertainty in our knowledge of neutrino masses and mixing angles will decrease 
considerably.
Many of the proposed models of neutrino mass and mixing will be tested. 
The Maki-Nakagawa-Sakata-Pontecorvo 
(MNSP) neutrino mixing matrix 
describes 
the unitary transformation between the mass  
and  flavor bases of the neutrinos. In vacuum it is given by 
$U_{\alpha j}=\la\alpha|j\ra$,
with $\alpha = e,\,\mu,\,\tau$ and $j=1,\,2,\,3$, i.e.~rows are 
labeled from top to bottom by the flavor indices,
and columns are labeled left to right by mass-eigenstate indices. 

Three mixing angles and a phase comprise the conventional parametrization of 
the vacuum mixing matrix, as established by the Particle Data Group (PDG)~\cite{PDG}:
{\small 
\bea {vacPDG}  
\hspace{-.5cm}U 
 &=& R_{32}(\theta_{32})\,U^{\dagger}_{\delta}\,
   R_{13}(\theta_{13})\,U_{\delta}\,R_{21}(\theta_{21}) \\ 
 &=&
\hspace{-0.2cm}\left( \bad 
c_{21}  c_{13} & s_{21}  c_{13} & s_{13}  e^{-i \delta}  \\ 
-s_{21}  c_{32} - c_{21}  s_{32}  s_{13}  e^{i \delta} 
& c_{21}  c_{32} - s_{21}  s_{32} s_{13}  e^{i \delta} 
& s_{32}  c_{13}  \\ 
s_{21}  s_{32} - c_{21}  c_{32}  s_{13}  e^{i \delta} & 
- c_{21}  s_{32} - s_{21}  c_{32}  s_{13}  e^{i \delta} 
& c_{32} c_{13}  \nonumber\\ 
\ea   
\right)
\eea
}
\noindent
where $R_{jk}(\theta_{jk})$ describes a rotation in the $jk$-plane
through angle $\theta_{jk}$, $U_{\delta}  = {\rm diag}(e^{i\delta/2},\,1,\,e^{-i\delta/2})$,
and $s_{jk} = \sin \theta_{jk}$, $c_{jk} = \cos \theta_{jk}$. 
%
We have omitted two additional Majorana phases,
as they do not affect neutrino oscillations. 

From Eq.~\rf{vacPDG}, one gleans that the PDG mixing angles are related to certain 
observable moduli of matrix elements as 
\begin{eqnarray} \label{eq:defobs}
\sin^2 \theta_{13} &=& \hspace{0.3cm} |U_{e3}|^2~~,~~ \\
\sin^2 \theta_{21} &=& {\D |U_{e2}|^2}   / ({\D 1 - |U_{e3}|^2})~~,~~\\
\sin^2 \theta_{32} &=& {\D |U_{\mu 3}|^2}/ ({\D 1 - |U_{e3}|^2})~.
\end{eqnarray}
In the order listed, these important moduli are inferred from 
terrestrial (long-baseline or reactor) data, solar data, and atmospheric data.
Finally, the CP-invariant of Jarlskog is given by
\be \label{eq:defjcp}
J_{\rm CP} = -{\rm Im}\{ U_{e 1} \, U_{\mu 3} \, U_{e 3}^\ast \, 
U_{\mu 1}^\ast \}\,.
\ee
In terms of the PDG parametrization in Eq.~\rf{vacPDG} this is  \\
$J_{\rm CP} = \frac 18 \, \sin 2 \theta_{21} \, \sin 2 \theta_{32} \, 
\sin 2 \theta_{13} \, \cos \theta_{13} \, \sin \delta$. 
Global 3-flavor fits to data give 
the following (1$\sigma$) and 3$\sigma$ ranges for the PDG mixing angles 
\cite{data}:   
\begin{eqnarray} \label{eq:data}
\sin^2 \theta_{21}
 &=& 0.32\,(\pm 0.02)~^{+0.08}_{-0.06} ~, \nonumber \\
\sin^2\theta_{32} &=& 0.45\,(\pm 0.07)~^{+0.20}_{-0.13} ~,\\
\sin^2 \theta_{13} &<& (0.02)~0.050~.\nonumber
\end{eqnarray}

The central values of these inferred mixing angles 
are quite consistent with the TriBiMaximal values~\cite{tribi},
given implicitly by 
$\sin^2 \theta_{21} = \frac 13$, $\sin^2 \theta_{32} = \frac 12$ 
and  $\sin^2 \theta_{13} = 0$. 
The resulting angles are
$\theta_{32}= \frac{\pi}{4} {\rm rad} = 45^\circ$, 
$\theta_{21}=
\sin^{-1} \sqrt{1/3} = 0.6155\ldots {\rm rad}
=35.2644\dots^\circ$, and $\theta_{13}=0$. 
%
%
Explicitly, the TriBiMaximal mixing matrix is
{\small
\beq{tribimax}
\Utbm = R_{32}\left(\frac\pi 4\right)R_{21}\left(\sin^{-1}\frac{1}{\sqrt{3}}\right) 
 =  
%
%
\frac{1}{\sqrt 6}
\left(
\begin{array}{rrc}
 2 &  \sqrt{2} &  0                  \\
-1 &  \sqrt{2} &  \sqrt{3} \\
 1 &  -\sqrt{2} & \sqrt{3}
\end{array}
\right).
\eeq
}
%
%
\noindent
TriBiMaximal is a good zeroth order approximation to reality.
However, we expect  that even if some flavor symmetry is embedded in Nature which 
leads to zeroth order TriBiMaximal mixing,
in general there will be  deviations from this scheme~(see for example, Ref.~\cite{devs}). 

In this Letter, we present and develop a 
parametrization of the MNSP matrix which is completely general, 
but has the TriBiMaximal matrix as its 
zeroth order basis.
We call the parametrization ``TriMinimal''. 
To accommodate the four independent parameters in $U$, we introduce as three 
small quantities $\epsilon_{jk}$, $jk = 21,32,13$,  
the deviation of the $\theta_{jk}$ from their TriBiMaximal values; 
and we retain the usual CP-violating phase $\delta$. 
TriBiMaximal mixing, given in Eq.~(\ref{tribimax}),
is recovered in the limit of all three $\epsilon_{jk}=0$.   
In other parametrizations, all three small parameters are typically coupled in 
the description of each $\theta_{jk}$~\cite{BHS}. 
We illustrate the utility of the TriMinimal parametrization by deriving a 
rather simple result for the flavor evolution of neutrinos traveling large distances.
This includes atmospheric and astrophysical neutrinos.
Our new result is presented as an expansion to second order in the three 
small $\eps_{jk}$.


The TriMinimal parametrization is given by 
\bea{Ualt}
U_{\rm TMin} &=& 
R_{32}\left(\frac{\pi}{4}\right)
\,U_\eps(\eps_{32};\eps_{13},\delta;\eps_{21})
\,R_{21}\left(\sin^{-1}\frac{1}{\sqrt{3}}\right)
\nonumber\\
 &=&
\left(
\begin{array}{rrr}
 \sqrt{2} & 0 & 0 \\
 0 &  1 & 1 \\ 
 0 & -1 & 1 
\end{array}
\right)\;\;
\frac{\Ueps^{\rm }}{\sqrt 6}\;\;
\left(
\begin{array}{rrr}
 \sqrt{2} & 1 & 0 \\
-1 & \sqrt{2} & 0 \\ 
 0 & 0 & \sqrt{3}
\end{array}
\right)\,, 
\nonumber\\
%
\mbox{with }
%
\Ueps^{\rm } &=& R_{32}(\eps_{32})\,U_\delta^\dag\,R_{13}(\eps_{13})\,U_\delta\,R_{21}(\eps_{21})
\eea
chosen to have just the form of the PDG parametrization.
And just as in the PDG parametrization, $U_\eps$ is unitary,
and therefore so is TriMinimal $U_{\rm TMin}$.
The simplicity of Eq.~\rf{Ualt}
is a fortuitous result 
of the fact that it is the middle rotation angle ($\theta_{13}$) 
in the PDG parametrization that 
is set identically to zero in the TriBiMaximal scheme.
%
%

From Eq.~\rf{Ualt} and Eqs.~(\ref{eq:defobs}--\ref{eq:defjcp}), 
one obtains the neutrino mixing observables in terms of the 
TriMinimal parameters.  
The exact result and the expansion up to 
second order in the $\eps_{jk}$ are: 
%
\begin{eqnarray} \label{eq:obsII} \D 
\sin^2 \theta_{21} &=& \frac 13 \left( \cos \epsilon_{21} + 
\sqrt{2} \, \sin \epsilon_{21} \right)^2  \\ & \simeq  & \nonumber 
\frac 13 + \frac{2\sqrt{2}}{3} \, \epsilon_{21} 
 + \frac{1}{3} \, \epsilon_{21}^2 ~,\\
\D
\sin^2 \theta_{32} &=& \frac 12 + \sin \epsilon_{32} \, \cos \epsilon_{32} 
\simeq  \frac 12 + \epsilon_{32} ~,\\[0.2cm]\D
U_{e3} &=& \sin \epsilon_{13} \, e^{-i \delta}   ~.\\
 && \nonumber
\eea
One sees in the above expressions that the TriMinimal parametrization maintains 
the simple parametrization of $U_{e3}$.
This is inevitable, for $\eps_{13}$ and $\delta$ are just the standard PDG
parameters $\theta_{13}$ and $\delta$.

Being a 3-flavor quantity, 
$J_{\rm CP}$ depends on all three $\epsilon_{jk}$'s.
It's expansion is
\begin{eqnarray} \label{eq:J2}
\D
J_{\rm CP} &=& \frac{\sin \delta}{24} \, 
\cos 2 \epsilon_{32} \, \sin 2 \epsilon_{13} 
\, \cos \epsilon_{13} \\ \nonumber 
& & \times \left(2\sqrt{2} \, \cos 2  \epsilon_{21} 
+  \sin 2\epsilon_{21}\right) \\ \nonumber 
& \simeq&   \frac{1}{3 \sqrt{2}}\,(\eps_{13}\sin \delta)\,
 (1+\frac{1}{\sqrt{2}}\,\epsilon_{21})\,.
\end{eqnarray}
%
At lowest order, $J_{\rm CP}$ depends on just $\eps_{13}\,\sin\delta$.
$J_{\rm CP}$ has no dependence on $\eps_{32}$ at first or second order. 

In the above formulae, we have truncated the expansions in powers of $\eps_{jk}$ at quadratic order,
since cubic order is likely to be immeasurably small.
The allowed ranges of the small $\eps_{jk}$ are obtained from the 
$(1\sigma)~3\sigma$ ranges 
of the large oscillation angles in Eq.~(\ref{eq:data}).
The allowed ranges are:
\begin{eqnarray} \label{eq:epsIIrange}
-0.08~(-0.04)\le &\epsilon_{21}& \le (0.01)~0.07 ~,\\ 
-0.18~(-0.10)\le &\epsilon_{32}& \le (0.04)~0.15 ~,\\ \label{eq:epsIIrangea} 
&|\epsilon_{13}|& \le (0.14)~0.23~,
\end{eqnarray}
while the CP-invariant lies in the range $|J_{\rm CP}| \le (0.03)~0.05$. 

Let us emphasize two virtues of TriMinimality.
With the ordering of the (small-angle) rotations in Eq.~(\ref{Ualt}) 
chosen in accord with the PDG parametrization, \\
%
(i) each  $\epsilon_{jk}$ is directly interpretable as the deviation of 
the associated $\theta_{32}$, $\theta_{13}$, or $\theta_{21}$ from its  
TriBiMaximal value; due to the non-commutivity of rotation matrices,
this feature is not shared with other parametrizations of the MNSP matrix~\cite{BHS},
but rather is unique to the TriMinimal parametrization. \\
%
(ii) the usual PDG result for $U_{e3}$, namely, 
$U_{e3}=\sin\eps_{13}\,e^{-i\delta}$, maintained.
%

From Eq.~\rf{Ualt}, 
it is straightforward to derive the expansion of $U_{\rm TMin}$ in powers of the 
three $\eps_{jk}$ and single $\delta$~\cite{LW07,note1}.
%
%
%
%
The result is
\begin{widetext}
\bea{U2ndII}
U_{\rm TMin}
= U_{\rm TBM} 
&-&
 \frac{\eps_{21}}{\sqrt{6}}
\left(
\ba{rrr}
\sqrt{2} & -2 & 0 \\
\sqrt{2} & 1 & 0 \\
- \sqrt{2} & -1 & 0 
\ea
\right)
-\frac{\eps_{32}}{\sqrt{6}}
\left(
\ba{rrr}
0 & 0          & 0 \\
 -1 & \sqrt{2} & -\sqrt{3} \\
 -1 & \sqrt{2} &  \sqrt{3} 
\ea
\right)
-\frac{\eps_{13}}{\sqrt{6}}
\left(
\ba{ccc}
0 & 0        & -\sqrt{6}\,e^{-i\delta} \\
\sqrt{2}\,e^{i\delta} & e^{i\delta} & 0 \\
\sqrt{2}\,e^{i\delta} & e^{i\delta} & 0
\ea
\right) \\
 &-& 
\frac{\eps^2_{21}}{2\sqrt{6}}
\left(
\ba{rrr}
 2 & \sqrt{2}  & 0 \\
-1 & \sqrt{2}  & 0 \\
 1 & -\sqrt{2} & 0 
\ea
\right)
-\frac{\eps^2_{32}}{2\sqrt{6}}
\left(
\ba{rrc}
  0 & 0 & 0 \\
 -1 & \sqrt{2} & \sqrt{3} \\
  1 & -\sqrt{2} & \sqrt{3}
\ea
\right)
-\frac{\eps^2_{13}}{2\sqrt{6}}
\left(
\ba{rcc}
 2 & \sqrt{2} & 0 \\
 0 & 0 & \sqrt{3} \\ 
 0 & 0 & \sqrt{3}
\ea
\right) \nonumber\\
 &-&
\frac{\eps_{21}\,\eps_{32}}{\sqrt{6}}
\left(
\ba{crr}
  0 & 0 & 0 \\
 -\sqrt{2} & -1 & 0 \\
 -\sqrt{2} & -1 & 0
\ea
\right)
-\frac{\eps_{21}\,\eps_{13}\,e^{i\delta}}{\sqrt{6}}
\left(
\ba{rcr}
 0 & 0 & 0\\
 -1 & \sqrt{2} & 0 \\ 
 -1 & \sqrt{2} & 0
\ea
\right)
-\frac{\eps_{32}\,\eps_{13}\,e^{i\delta}}{\sqrt{6}}
\left(
\ba{rrc}
 0 & 0 & 0 \\
 \sqrt{2} & 1 & 0 \\ 
-\sqrt{2} & -1 & 0
\ea
\right)+{\cal O}(\eps^3)\,.\nonumber
\eea
\end{widetext}

In the remainder of this paper we present and explore the TriMinimal parametrization 
of the phase-averaged mixing that describes atmospheric and astrophysical 
neutrino flavor propagation.
We first introduce the matrix $\U2$ of classical probabilities, 
defined by 
$(\U2)_{\alpha j}\equiv |U_{\alpha j}|^2$.
The full matrix of squared elements, through order ${\cal O}(\eps^2)$, is
\begin{widetext}
\beq{2ndU2compactII}
\U2_{\rm TMin} = \frac{1}{6}
\left\{ \, 
\left(\begin{array}{ccc}
 4 & 2 & 0   \\
 1 & 2 & 3 \\
 1 & 2 & 3
\end{array}\right)
-E_1
\left(
\ba{rrr}
 0 & 0 & 0 \\
 -1 &  1 & 0 \\
  1 & -1 & 0
\ea
\right)
-E_2
\left(
\ba{rrr}
 2 & -2 & 0 \\
-1 & 1 & 0 \\
-1 & 1 & 0 
\ea
\right)
- 2\,\eps_{32}
\left(
\ba{rrr}
0 & 0  & 0 \\
1 & 2 & -3 \\
-1 & -2 & 3 
\ea
\right)
-\eps^2_{13}
\left(
\ba{crr}
 4 & 2 & -6 \\
 -2 & -1 & 3 \\
 -2 & -1 & 3
\ea
\right)
\ \right\}\,,
%
\eeq
\end{widetext}
where $E_1=2\sqrt{2}\,\eps_{13}\cos\delta+2\,\eps_{21}\,(\eps_{13}\,\cos\delta-2\sqrt{2}\,\eps_{32})$,
and $E_2=2\sqrt{2}\,\eps_{21}+\eps^2_{21}$.
That there are four independent terms in \rf{2ndU2compactII}
reflects the fact that there are four independent moduli in the neutrino mixing matrix~\cite{note2}.

Some useful results follow immediately from this matrix.
For example, in models where neutrinos are unstable,
only the lightest neutrino mass-eigenstate arrives at Earth 
from cosmically-distant sources~\cite{unstable}.
Flavor ratios at Earth for the normal mass-hierarchy are
$\U2_{e1} : \U2_{\mu 1} : \U2_{\tau 1}$,
and for the inverted mass-hierarchy are
$\U2_{e3} : \U2_{\mu 3} : \U2_{\tau 3}$.
These ratios may be read off directly from the 1st and 3rd columns of Eq.~\rf{2ndU2compactII}.
As another example, emanating solar neutrinos are nearly pure $\nu_2$ mass states~\cite{nu2};
consequently, their flavor ratios at Earth are mainly given by the 2nd column of~\rf{2ndU2compactII}.

As is well-known, 
the neutrino mixing probabilities for phase-averaged propagation 
(appropriate when the oscillation phase $(\Delta m^2 L/4E)$ is much larger than one) 
are given by   
\begin{equation} \label{eq:Pab}
P_{\nu_\alpha\leftrightarrow \nu_\beta} = 
\sum\limits_i |U_{\alpha i}|^2 \,|U_{\beta i}|^2
= \U2\U2^T\,.
\end{equation}
The full result in terms of the $\eps_{jk}$ is
\begin{widetext}
\beq{UUcompact}
(\U2^{\rm }\U2^T )_{\rm TMin}
= 
\frac{1}{18}
\left\{
\left(
\begin{array}{ccc}
 10 & 4 & 4 \\
 4 & 7 & 7  \\
 4 & 7 & 7  
\end{array}
\right)
+ A\,
\left(
\ba{rrr}
 4 & -2 & -2 \\
-2 & 1 & 1 \\
-2 & 1 & 1 
\ea
\right) 
%
+
B\,
\left(
\ba{rrr}
 0 & 1 & -1 \\
 1 & -1 & 0 \\
-1 & 0 & 1
\ea
\right) 
%
+ 
C\,
\left(
\ba{rrr}
  0 & 0 & 0 \\
  0 & 1 & -1 \\
  0 & -1 & 1
\ea
\right)
\right\}
+{\cal O}(\eps^3)\,, \\ 
%
\eeq
where
%
%
%
%
\beq{A}\nonumber
A(\eps_{21}; \eps_{21}^2,\eps_{13}^2) = 
   -(2\sqrt{2}\,\eps_{21}-7\eps^2_{21}+5\,\eps^2_{13})\,,
\eeq
\beq{B} \nonumber
B(\eps_{32}, \eps_{13}\cos\delta; \eps_{21}\eps_{32}, \eps_{21}\eps_{13}\cos\delta) =
   -2\,(4\,\eps_{32}-\sqrt{2}\,\eps_{13}\cos\delta
        +4\sqrt{2}\,\eps_{21}\,\eps_{32}+7\,\eps_{21}\,\eps_{13}\cos\delta)\,,
\eeq
\beq{C} \nonumber
C(\eps_{32}^2, \eps_{32}\eps_{13}\cos\delta, (\eps_{13}\cos\delta)^2) =
  4\,(7\,\eps^2_{32}+2\,(\eps_{13}\cos\delta)^2
   +\sqrt{2}\,\eps_{32}\,\eps_{13}\cos\delta)\,.
\eeq
%
The symmetric matrix in Eq.~\rf{UUcompact} contains the six explicit flavor mixing probabilities
\bea{Ps}
P_{\nu_e\leftrightarrow\nu_e} &=& \frac{1}{18}(10+4\,A)\,,\quad
P_{\nu_\mu \leftrightarrow\nu_\mu} = \frac{1}{18}(7+A-B+C)\,,\quad
P_{\nu_\tau \leftrightarrow\nu_\tau} = \frac{1}{18}(7+A+B+C)\,,\nonumber\\
P_{\nu_e \leftrightarrow\nu_\mu} &=& \frac{1}{18}(4-2\,A+B)\,,\quad
P_{\nu_e \leftrightarrow\nu_\tau} = \frac{1}{18}(4-2\,A-B)\,,\quad
P_{\nu_\mu \leftrightarrow\nu_\tau} = \frac{1}{18}(7+A-C)\,.
\eea
\end{widetext}

At first sight, it seems remarkable that only three terms, $A$, $B$, and $C$, have emerged to 
parametrize the six elements $P_{\nua\leftrightarrow\nub}$ in $\U2\U2^T$.
However, this is inevitable, for there are only three independent $P_{\nua\leftrightarrow\nub}$
as a result of the unitary sum rules $\sum_\beta P_{\nua\leftrightarrow\nub}=1$.

Notice that since each row in $({\underline U}\,{\underline U^T})$ 
partitions a flavor neutrino among all possible flavors,
each row must sum to unity at zeroth order in $\epsilon_{jk}$,
and to zero at each nonzero order in  $\epsilon_{jk}$ and $\cos\delta$.
Then, because of T-reversal invariance, or equivalently,
because $({\underline U}\,{\underline U^T})$
is a symmetric matrix, each column must also sum to unity at zeroth order in 
$\epsilon_{jk}$,
and sum to zero at each nonzero order in  $\epsilon_{jk}$ and $\cos\delta$.

$A$ and $B$ are of indeterminate sign, whereas C,
which may be written as 
$(2\sqrt{2}\,\eps_{13}\cos\delta+\eps_{32})^2+27\,\eps_{32}^2$, 
is manifestly positive semidefinite.
%
%
$A$ and $B$ contain terms linear in $\eps$'s, as well as quadratic terms;
$C$ is purely quadratic in $\eps$'s.
For certain values of the $\eps_{jk}$, the second order corrections may dominate the first
order corrections. Consequently, $C$ should not be neglected~\cite{note3}.

%
%
%

%
%

The dependences of flavor oscillation and flavor mixing probabilities 
on first and second order corrections will be examined in considerable detail, 
for low and high energy neutrinos, in~\cite{PRW}.  
Here we present an interesting application of the phase-averaged mixing matrix 
in Eq.~\rf{UUcompact}. 
The most common source for atmospheric and astrophysical neutrinos is thought to be 
pion production and decay.
The pion decay chain generates an initial neutrino flux with flavor composition 
given approximately~\cite{lipari} by 
$\Phi_e^0 : \Phi_\mu^0 : \Phi_\tau^0 = 1 : 2 : 0$ for the neutrino fluxes.
According to Eq.~\rf{UUcompact}, the fluxes $\Phi_\alpha$ arriving at earth 
have a flavor ratio of 
\begin{small}
\be \label{eq:flux120} \nonumber
\Phi_e : \Phi_\mu : \Phi_\tau =  
1+\frac{1}{18}(2\,B): 1-\frac{1}{18}(B-2\,C): 1-\frac{1}{18}(B+2\,C)\,.
\ee
\end{small}
Violation of $\numu\leftrightarrow\nutau$ symmetry, exact with TriBiMaximal mixing,
are directly assessed via the flavor ratio 
\beq{mu2tau}
\frac{\Phi_\mu}{\Phi_\tau} = 1+\frac{2}{9}C +{\cal O}(\eps^3)\,.
\eeq
From this result we may infer two lessons:
$C \ge 0$, so we learn that $\Phi_\mu\ge \Phi_\tau$ is an 
inevitable consequence for pion-produced, astrophysical neutrinos;
and we see an explicit example where the second order correction dominates
over the first order correction (exactly zero in this case).
The TriMinimal parametrization and expansion has made this result transparent.

In summary, we have presented the TriMinimal parametrization of the MNSP matrix. 
Three small parameters $\epsilon_{jk} \ll 1$, each equal to the  
the deviation of one of the measured quantities 
$\theta_{jk}$ from its TriBiMaximal value, plus the usual CP-violating phase $\delta$,
comprise the parametrization.
The TriMinimal parametrization leads to simple formulas for neutrino flavor mixing.
The proposed parametrization in Eq.~\rf{Ualt}, 
the expansions in~\rf{U2ndII}, \rf{2ndU2compactII}, and \rf{UUcompact}, 
and the mixing probabilities in~\rf{Ps} 
are the main results of this Letter.
Simple properties of the TriMinimal parametrization are not shared by other parametrizations.

\noindent
{\sl Note added}: 
As this paper was being written, a very similar proposal
for parametrization was posted on the ArXiv~\cite{King}.
There, the utility of TriMinimal parameters for terrestrial flavor oscillations was emphasized;
here, we emphasize the utility for phase-averaged atmospheric and astrophysical flavor mixing.
After this paper was written, we were made aware of some earlier but different parametrizations
of ``almost TriBiMaximal'' mixing matrices~\cite{earlier}.

\noindent
{\sl Acknowledgments}:
W.R.~was supported by the Deutsche Forschungsgemeinschaft 
in the Sonderforschungsbereich 
Transregio 27  and under project 
number RO--2516/3--2, as well as by the EU program ILIAS N6 ENTApP WP1. 
S.P.~and T.J.W.~thank M. Lindner and 
MPI-Heidelberg for support and hospitality, 
and acknowledge support from U.S.~DoE  
grants DE--FG03--91ER40833 and DE--FG05--85ER40226.


\begin{thebibliography}{99}
\bibitem{note1} 
In deriving this formula, it is useful to expand 
in powers of $\eps$ the following form of the 
individual rotation matrices:
%
\[
\hspace{0.6cm}
R(\eps)=\left(
\ba{rr}
\cos\eps & \sin\eps \\
-\sin\eps & \cos\eps 
\ea\right)
=
\unitmatrix-2\sin\frac{\eps}{2}\left(
\ba{rr}
\sin\frac{\eps}{2}& -\cos\frac{\eps}{2}\\
\cos\frac{\eps}{2}&\sin\frac{\eps}{2}
\ea\right).
\]
\end{thebibliography}
\end{document}